\documentclass[a4paper]{article}

\usepackage{INTERSPEECH_v2}

\title{Deep Speaker Feature Learning for Text-independent Speaker Verification}
\name{Lantian Li, Yixiang Chen, Ying Shi, Zhiyuan Tang, Dong Wang$^*$}
\address{
  Center for Speech and Language Technologies, Tsinghua University, China}
\email{\{lilt13,wangdong99\}@mails.tsinghua.edu.cn, \{chenxy,shiying,tangzy\}@cslt.riit.tsinghua.edu.cn}

\begin{document}

\maketitle
\begin{abstract}

  Recently deep neural networks (DNNs) have been used to learn speaker features.
  However, the quality of the learned features
  is not sufficiently good, so a complex back-end model, either
  neural or probabilistic, has to be used to address the residual uncertainty when
  applied to speaker verification, just as with raw features. This paper presents
  a convolutional time-delay deep neural network structure (CT-DNN) for speaker feature learning.
  Our experimental results on the \emph{Fisher} database demonstrated that this CT-DNN can
  produce high-quality speaker features: even with a single feature (0.3 seconds including the context),
  the EER can be as low as 7.68\%. This effectively confirmed that the speaker trait
  is largely a deterministic short-time property rather than a long-time distributional pattern,
  and therefore can be extracted from just dozens of frames.

\end{abstract}
\noindent\textbf{Index Terms}: deep neural networks, speaker verification, speaker feature

\section{Introduction}
\label{sec:intro}

Automatic speaker verification (ASV) is an important biometric authentication technology.
As in most machine learning tasks, a key challenge of ASV is the intermixing of multiple variability
factors involved in the speech signal, which leads to great uncertainty when making genuine/imposter decision.
In principle, two methods can be employed to address the uncertainty: either
by extracting more powerful features which are sensitive to speaker traits but invariant to other
variations, or by constructing a statistical model that can describe the uncertainty and promote the
speaker factor.

Most of existing successful ASV approaches are model-based. For example, the
famous Gaussian mixture model-universal background model (GMM-UBM) framework~\cite{Reynolds00}
and the subsequent subspace models, including the joint factor analysis approach~\cite{Kenny07}
and the i-vector model~\cite{dehak2011front}. They are generative models and heavily utilize
unsupervised learning. Improvements have been achieved in two directions.
The first is to use a discriminative model to boost the discriminant
for speakers, e.g., the SVM model for the GMM-UBM approach~\cite{Campbell06} and the PLDA model for the
i-vector approach\cite{Ioffe06}.
The second is to use supervised learning to produce a better representation for the acoustic space. For
example, the DNN-based i-vector method~\cite{Kenny14,lei2014novel}.
Almost all the model-based methods use raw features, e.g., the popular Mel frequency cepstral coefficients (MFCC)
feature.


In spite of the predominant success of the model-based approach, researchers never stop searching for
`fundamental' features for speaker traits. The motivation is two-fold: from the engineering perspective,
if a better feature is found, the present complex statistical models can be largely discarded;
and from the cognitive perspective, a fundamental
feature will help us understand how speaker traits are embedded in speech signals. Driven by these
motivations, many researchers put their effort in `feature engineering' in the past several decades,
and new features were proposed occasionally, from perspectives of different knowledge domains~\cite{Kinnunen10}.
However, compared to the remarkable achievement with the model-based approach, the reward from the feature engineering
is rather marginal. After decades, we find the most useful feature in our hand is still MFCC. Interestingly,
the same story was also told in other fields of speech processing, particularly in automatic speech recognition (ASR),
until very recently after deep learning involved.

The development of deep learning changed the story.
Different from the historic feature engineering methods that design features by human
knowledge, deep learning can \emph{learn} features automatically from vast raw data, usually by
a multi-layer structure, e.g., a deep neural network (DNN). By the layer-by-layer processing,
task-related information can be preserved and strengthened, while task-irrelevant variations
are diminished and removed. This feature learning has been
demonstrated to be very successful in ASR, where the learned features have shown to be highly representative
for linguistic content and very robust against variations of other factors~\cite{hinton2012deep}.

This success of feature learning in ASR has motivated researchers in ASV to learn
speaker sensitive features. The primary success was reported by Ehsan et al. on a text-dependent
task~\cite{ehsan14}. They constructed a DNN model with $496$ speakers
in the training set as the targets. The frame-level features were read from the activations of the last hidden layer,
and the utterance-level representations (called `d-vector') were obtained by averaging over frame-level
features. In evaluation, the decision score was computed as a simple cosine distance between the d-vectors of the
enrollment utterance(s) and the test utterance. The authors reported worse performance with the d-vector system
compared to the conventional i-vector baseline, but after combining the two systems, better
performance was obtained. This method was further extended by a number of researchers.
For example, Heigold et al.~\cite{heigold2016end} used an LSTM-RNN to learn utterance-level representations directly
and reported better performance than the i-vector system on the same text-dependent task
when a large database was used (more than $4,000$ speakers). Zhang et al.~\cite{zhang2017end}
utilized convolutional neural networks (CNN) to learn speaker features and an attention-based model to
learn how to make decisions, again on a text-dependent task. Liu et al.~\cite{liu2015deep} used the DNN-learned
features to build the conventional i-vector system.
Recently, Snydern et al.~\cite{snyderdeep16} migrated the DNN-based approach to text-independent tasks, and
reported better performance than the i-vector system when the training data is sufficiently large (102k speakers).
All these following-up studies, however, are not purely feature learning: they all involve a complex back-end
model, either neural or probabilistic, to gain reasonable performance. This is perfectly fine from
the perspective of both research and engineering, but departs from the initial goal of feature learning:
we hope to discover a feature that is sufficiently general and discriminative so that it can
be employed in a broad range of applications without heavy back-end models. This has been achieved
in ASR, but not in speaker verification yet.


In this paper, we present a simple but effective DNN structure that involves two convolutional layers and
two time-delayed full-connection layers to learn speaker features.
Our experiments demonstrated that this simple model can learn very strong speaker sensitive features, using
speech data of only a few thousand of speakers. The learned feature does not require complex back-end models:
a simple frame averaging is sufficient to produce a strong utterance-level speaker vector, by which a
simple cosine distance is good enough to perform text-independent ASV tasks. These results actually
demonstrated that it is possible to discover speaker information from a short-time speech segment ($300$ ms),
by only a couple of simple neural propagation.

The rest of this paper is organized as follows. Section~\ref{sec:rel} describes the
related work, and Section~\ref{sec:theory} presents the new DNN structure.
The experiments are presented in Section~\ref{sec:exp}, and Section~\ref{sec:conl}
concludes the paper.

\section{Related work}
\label{sec:rel}

Our work is a direct extension of the d-vector model presented by Ehsan et al~\cite{ehsan14}.
The extension is two-fold: a CNN/TDNN structure that emphasizes on
temporal-frequency filtering, more resemble to the traditional feature engineering; an experiment on
a text-independent task demonstrated that the learned feature is independent of linguistic content
and highly speaker sensitive.

This work is different from most of the existing neural-based ASV methods.
For example, the RNN-based utterance-level representation learning~\cite{heigold2016end} is
attractive, but the RNN pooling shifts the focus to the entire sentence, rather
than frame-level feature learning. The end-to-end neural models proposed by Snyder~\cite{snyderdeep16}
and Zhang~\cite{zhang2017end} both involve a back-end classifier, which weakens the feature learning component: it is unknown
whether the speaker-discriminant information is learned by the classifier or by the feature
extractor. Therefore, the features are not necessarily speaker discriminative and
are less generalizable, as the feature extractor depends on the classifier.

This work is also different from the methods that combine DNN features and
statistical models. In these methods, some speaker information is learned
in the feature, but not sufficient. Therefore, the
feature is still primary and thus
a statistical model has to be used to address the inherent uncertainty.
For example, Liu et al.~\cite{liu2015deep} used an ASR-ASV
multi-task DNN to produce frame-level features and substituted them for MFCC
to construct GMM-UBM and i-vector systems. Yao et al.~\cite{yao2016speaker}
proposed a similar approach, though they used ASR-oriented features to train
GMMs for splitting the acoustic space, and the original ASV-oriented features as the acoustic
feature to construct the i-vector model.

An implication behind the above methods is that the speaker feature learning
is still imperfect: the speaker traits have not
been fully extracted and other irrelevant variations still exist,
and therefore some back-end models have to be utilized to improve
the discriminative power.
In this paper, we will show that a better network design can significantly
improve the quality of the feature learning, hence greatly alleviating the
reliance on a back-end model.

\section{CT-DNN for feature learning}
\label{sec:theory}

This section presents our DNN structure for speaker sensitive feature learning.
This structure is an extension to the model proposed in~\cite{ehsan14}, by using
convolutional layers to extract local discriminative patterns from the temporal-frequency space,
and time-delayed layers to increase the effective temporal context for each frame.
We call this structure as CT-DNN.


    \begin{figure*}[htb]
    \centering
    \includegraphics[width=1\linewidth]{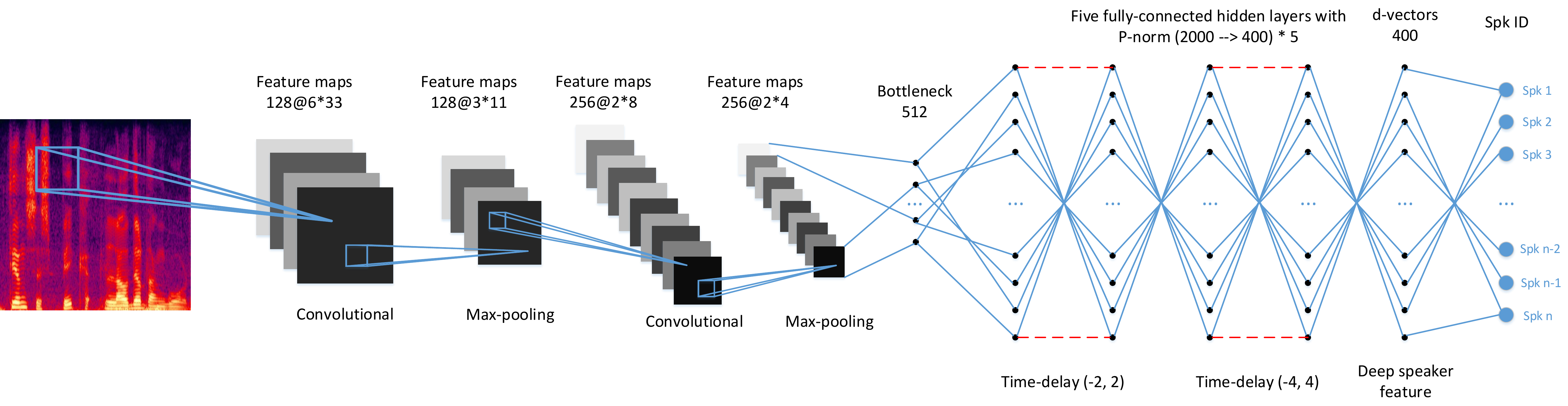}
    \caption{The CT-DNN structure used for deep speaker feature learning.}
    \label{fig:ctdnn}
    \vspace{-2mm}
    \end{figure*}

Figure~\ref{fig:ctdnn} illustrates the CT-DNN structure used in this work.
It consists of a convolutional (CN) component and a time-delay (TD) component, connected
by a bottleneck layer consisting of $512$ hidden units.
The convolutional component involves two CN layers, each followed by a max pooling.
This component is used to learn local patterns that are useful in representing speaker traits.
The TD component involves two TD layers, each followed by a P-norm layer.
This component is used to extend the temporal context.
The settings for the two components, including the patch size, the number of feature maps, the
time-delay window, the group size of the P-norm, have been shown in Figure~\ref{fig:ctdnn}.
A simple calculation shows that with these settings, the size of the effective
context window is $20$ frames.
The output of the P-norm layer is projected to a feature layer consisting of $400$ units,
which is connected to the output layer whose units correspond to
the speakers in the training data.

This CT-DNN model is rather simple and can be easily trained. In our study, the natural stochastic gradient
descent (NSGD)~\cite{povey2014parallel} algorithm was employed to conduct the optimization.
Once the DNN model has been trained, the speaker feature can be read from the feature layer,
i.e., the last hidden layer of the model. As in~\cite{ehsan14}, the utterance-level representation
of a speech segment can be simply derived by averaging the speaker features of all the frames
of the speech segment.

Following the name convention of the previous work~\cite{ehsan14,li2015improved}, the utterance-level
representations derived from the CT-DNN are called d-vectors.
During test, the d-vectors of the test and enrollment utterances are produced respectively.
A simple cosine distance between these two vectors can be computed and used
as the decision score for the ASV task. Similar with i-vectors, some simple normalization
methods can be employed, such as linear discriminant analysis (LDA) and probabilistic LDA (PLDA).

\section{Experiments}
\label{sec:exp}

In this section, we first present the database used in the experiments, and then report the results with
the i-vector and the d-vector systems. All the experiments were conducted with the Kaldi toolkit~\cite{povey2011kaldi}.

\subsection{Database}
\label{sec:data}

The \emph{Fisher} database was used in our experiments. The training data and the test data are presented as follows.

\begin{itemize}
    \item \textbf{Training set}: It consists of $2,500$ male and $2,500$ female speakers, with $95,167$ utterances randomly selected from the \emph{Fisher} database, and each speaker has about $120$ seconds speech segments.
        This dataset was used for training the UBM, the T-matrix, and the LDA/PLDA models
        of the i-vector system, and the CT-DNN model of the d-vector system.
    \item \textbf{Evaluation set}: It consists of $500$ male and $500$ female speakers randomly selected from
        the \emph{Fisher} database. There is no overlap between the speakers of the training set and the evaluation set.
        For each speaker, $10$ utterances are used for enrollment and the rest for test.

\end{itemize}

The test were conducted in $4$ conditions, each with a different setting in the length of the enrollment and the test utterances.
The conditions are shown in Table~\ref{tab:data}. All the test conditions involve pooled male and female trials. Gender-dependent test
exhibited the same trend, so we just report the results with the pooled data.

    \begin{table}[htp]
    \begin{center}
        \caption{Data profile of the test conditions.}
        \vspace{-1mm}
        \label{tab:data}
          \begin{tabular}{|l|c|c|c|c|}
           \hline
               Test condition    &   C(30-3) &    C(30-9)   &   C(30-18)   &    C(30-30)  \\
           \hline
           \hline
               Enroll. utts      &   10k     &    10k       &    10k       &     10k     \\
               Test utts         &   73k     &    24k       &    12k       &      7k   \\
           \hline
                Enroll. length    &    30s    &    30s       &     30s       &     30s   \\
                Test length      &    3s     &    9s        &     18s       &     30s   \\
           \hline
                Target trials    &   73k     &    24k       &     12k       &      7k  \\
                Nontarget trials &   36M     &    12M       &     5.9M      &     3.5M   \\
           \hline
          \end{tabular}
    \end{center}
    \vspace{-5mm}
    \end{table}

\subsection{Model settings}


We built an i-vector system as the baseline.
The raw feature involves $19$-dimensional MFCCs plus the log energy.
This raw feature is augmented by its first and second order derivatives, resulting in a
60-dimensional feature vector. This MFCC feature was used by the i-vector model.
The UBM was composed of $2,048$ Gaussian components, and the dimensionality of
the i-vector space was $400$. The dimensionality of the LDA projection space
was set to $150$.
The entire system was trained using the Kaldi SRE08 recipe.

For the d-vector system, the architecture was based on Figure~\ref{fig:ctdnn}.
The input feature was 40-dimensional Fbanks, with a symmetric $4$-frame window to
splice the neighboring frames, resulting in $9$ frames in total.
The number of output units was $5,000$, corresponding to the number of speakers in
the training data.
The speaker features were extracted from the last hidden layer (the feature layer in Figure~\ref{fig:ctdnn}),
and the utterance-level d-vectors were derived by averaging the frame-level features.
The transform/scoring methods used for the i-vector system were also used for the
d-vector system during the test,
including cosine distance, LDA and PLDA. The Kaldi recipe to reproduce our results
has been published online\footnote{http://data.cslt.org}.

\subsection{Main results}

The results of the i-vector system and the d-vector system in terms of equal error rate (EER\%)
are reported in Table~\ref{tab:baseline}. The two systems were trained with the entire training set,
and the results are reported for different conditions.

It can be observed that for both the two systems, improving the length of the test utterances
always improves performance. However, it seems that the performance improvement for the i-vector
system is more significant than the d-vector system. This is understandable as the i-vector system
relies on the statistical pattern of the features to build speaker vectors, so more speech frames will help.
In contrast, the d-vector system uses a simple average of the features to represent a speaker, so the contribution of
more speech frames is marginal.

The most interesting observation is the clear advantage
of the d-vector system in the C(30-3) condition, where the test utterances are
short. Because the d-vector system does not use
any powerful back-end model, this advantage on short utterances implies that the features learned by the CT-DNN model is
rather powerful. The advantage of neural-based models on short utterances
has been partially observed from DET curves by Snyder et al.~\cite{snyderdeep16},
and our results give a more clear evidence to this trend.

Another observation is that the LDA approach improves the d-vector system, while the PLDA approach
does not. The contribution of LDA suggests that there is still some non-speaker variation
within the learned feature, which needs more investigation. The failure of PLDA
with d-vectors is also a known issue in our previous work~\cite{li2015improved}. A possible
reason is that the residual noise within d-vectors is not Gaussian, and so
cannot be well modeled by the PLDA model. Again, more investigation is under going.

    \begin{table}[htb]
    \begin{center}
      \caption{The EER(\%) results with the i-vector and d-vector systems.}
       \vspace{-1mm}
      \label{tab:baseline}
          \begin{tabular}{|l|l|c|c|c|c|}
            \hline
            \multicolumn{2}{|c|}{}                 &\multicolumn{4}{c|}{EER\%}\\
            \hline
               Systems              &  Metric    &   C(30-3) &    C(30-9)   &   C(30-18)   &    C(30-30)\\
           \hline
           \hline
               i-vector             &    Cosine   &   3.77    &    1.54      &    0.80      &     0.49   \\
                                    &    LDA      &   3.11    &    1.24      &    0.82      &     0.59   \\
                                    &    PLDA     &   3.04    & \textbf{0.99}& \textbf{0.63} & \textbf{0.49}  \\
            \hline
               d-vector             &    Cosine   &   3.68    &    2.52      &    2.29      &     2.27    \\
                                    &    LDA      &\textbf{2.15} &   1.52    &    1.40      &     1.37    \\
                                    &    PLDA     &   6.90    &    3.54      &    2.95      &     2.74    \\
           \hline
          \end{tabular}
      \end{center}
      \vspace{-5mm}
   \end{table}

\subsection{Training data size}

\begin{figure}[htb]
    \centering
    \includegraphics[width=1\linewidth]{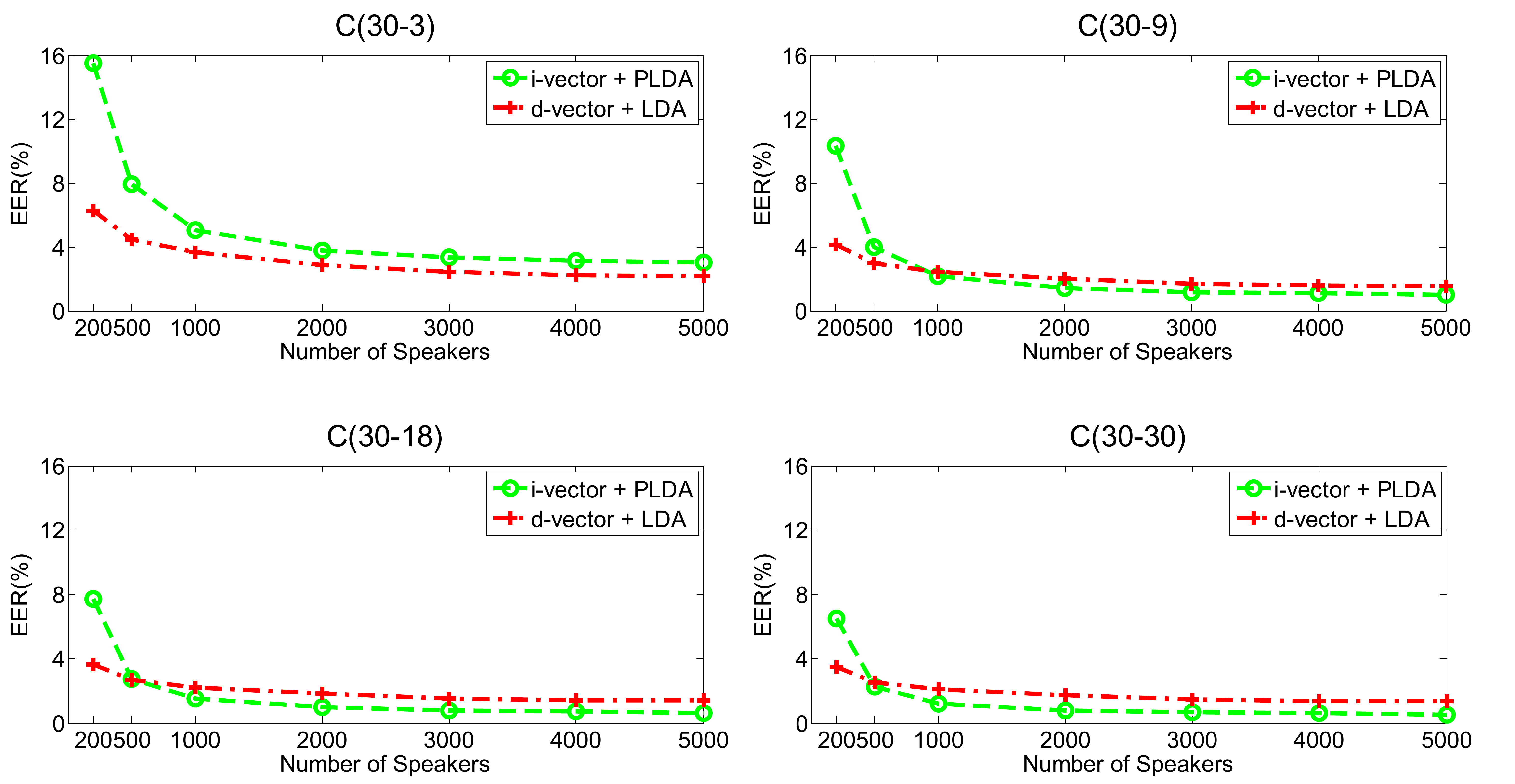}
    \caption{The EER(\%) results of the i-vector and d-vector systems trained with different sizes of training data. Each
    picture shows a particular test condition.}
    \label{fig:all}
    \vspace{-2mm}
    \end{figure}

  In order to investigate the data dependency of the feature learning approach, we
  change the size of the training data by selecting different numbers of speakers.
  The results of the best i-vector system (i-vector + PLDA) and the best d-vector system (d-vector + LDA)
  are reported in each of the four test conditions. The results are shown in Figure~\ref{fig:all},
  where each picture presents a particular test condition. It can be seen that
  in all the test conditions, for both the i-vector and d-vector systems,
  better performance is obtained with more training data, but the i-vector system seems
  benefit more from big data. This is a little different from our experience that
  deep neural models need more data than probabilistic models. It is also different
  from the observation in~\cite{snyderdeep16}, where the d-vector system obtained
  more performance improvement than the i-vector system when the number of speakers
  got very large (102k).

  We attribute the relatively less data sensitivity of the d-vector system
  to two factors: (1) the compact CN and TD layers in the CT-DNN
  structure require less training data; (2) the major restriction on performance
  of the d-vector system is not the model training, but the simple average back-end.
  In the C(30-3) condition where the test utterances are short, the impact of feature average
  is less significant, and so the true quality of the learned feature is exhibited,
  leading to a clear performance improvement when more training data are used.
  In other conditions, however, the improvement obtained by
  the big data training may have largely been masked by the average back-end.

\subsection{Feature discrimination}

\begin{figure}[htb]
    \centering
    \includegraphics[width=0.95\linewidth]{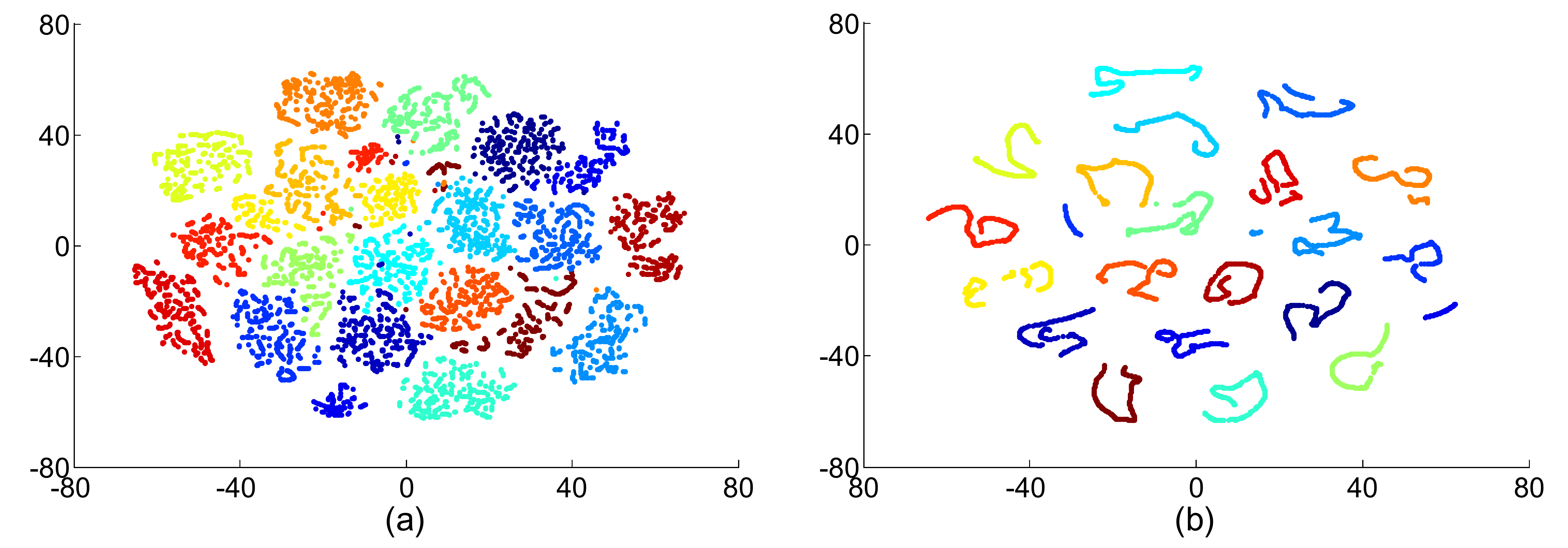}
    \caption{Deep features that are (a) randomly sampled (b) sequentially selected from an utterance. The
     picture is plotted by t-SNE, with each color representing a speaker. }
    \label{fig:tsne}
    \vspace{-2mm}
    \end{figure}

To check the quality of the learned speaker feature, we use t-SNE~\cite{saaten2008} to draw some feature samples from
$20$ speakers. The samples are selected in two ways: (a) randomly sample from all the speech frames
of the speaker; (b) choose a particular utterance. The results are presented in Figure~\ref{fig:tsne}.
It can be seen that the learned features are very discriminative for speakers, but there is still some
variation caused by linguistic content, as seen from the plot (b).

A more quantitative test for the feature quality is to examine the extreme case where the test speech is
only a few frames. Let's start from $20$ frames, which
is actually the effective context size of the CT-DNN, so only a single feature is produced.
More frames will produce  more features. Table~\ref{tab:ext1} and Table~\ref{tab:ext2} present
the results, where the length of the enrollment utterances is $30$ seconds and $3$ seconds, respectively.
The i-vector and d-vector models used in this experiment were trained with the entire training data.

The results with the d-vector system are striking: with only one feature ($20$ frames, or equally $0.3$ ms),
the EER can reach $13.54\%$ if the enrollment speech is only $3$ seconds, and $7.68\%$ when the enrollment speech is $30$ seconds!
In comparison, the i-vector system largely fails in these conditions.

These results demonstrated
that the CT-DNN model has learned a highly discriminative feature. From another perspective, these results
also demonstrated that the speaker trait is largely a deterministic short-time property rather than a
long-time distributional pattern, and therefore can be extracted from just dozens of speech frames.
This indicates that feature learning may be a more reasonable approach than the existing model-based approaches
that rely on statistical patterns of raw features.

    \begin{table}[htb]
    \begin{center}
      \caption{The EER(\%) results of the i-vector and d-vector systems with very short test utterances. The
      length of the enrollment utterances is $30$ seconds in average.}
      \vspace{-1mm}
      \label{tab:ext1}
          \begin{tabular}{|l|l|c|c|c|c|}
            \hline
               Systems              &    Metric   & 20 frames &   50 frames  &  100 frames   \\
           \hline
           \hline
               i-vector             &    Cosine   &   30.01    &    18.23    &    11.14      \\
                                    &    LDA      &   29.47    &    15.96    &     8.64      \\
                                    &    PLDA     &   29.29    &    15.71    &     8.34 \\
            \hline
               d-vector             &    Cosine   &\textbf{7.68} &  6.67      &    4.61      \\
                                    &    LDA      &   7.88     &\textbf{4.72} &\textbf{3.02}      \\
                                    &    PLDA     &  20.81     &    15.02     &    8.98    \\
           \hline
          \end{tabular}
      \end{center}
      \vspace{-6mm}
   \end{table}

    \begin{table}[htb]
    \begin{center}
      \caption{The EER(\%) results of the i-vector and d-vector systems with very short test utterances. The
      length of the enrollment utterances is $3$ seconds in average.}
      \vspace{-1mm}
      \label{tab:ext2}
          \begin{tabular}{|l|l|c|c|c|c|}
            \hline
               Systems              &    Metric   & 20 frames &   50 frames  &  100 frames   \\
           \hline
           \hline
               i-vector             &    Cosine   &   38.07    &    31.33     &   26.38    \\
                                    &    LDA      &   34.62    &    24.95     &   18.77    \\
                                    &    PLDA     &   33.67    &    21.97     &   15.08    \\
            \hline
               d-vector             &    Cosine   &\textbf{13.54} &  12.54    &   10.89      \\
                                    &    LDA      &   14.43   &\textbf{10.82} & \textbf{8.80}      \\
                                    &    PLDA     &  21.78     &    17.88     &   14.14    \\
           \hline
          \end{tabular}
      \end{center}
      \vspace{-5mm}
   \end{table}

\section{Conclusions}
  \label{sec:conl}

 This paper presented a CT-DNN model to learn speaker sensitive features. Our experiments
 showed that the learned feature is highly discriminative and can be used to achieve impressive
 performance when the test utterances are short. This result has far-reaching implications for
 both research and engineering: on one hand, it means that the speaker trait is a kind of
 short-time pattern and so should be extracted by short-time analysis (including neural learning)
 rather than long-time probabilistic modeling; on the other hand, our result suggests that
 the feature learning approach could/should be used when the test utterances are short,
 a condition that many practical applications are interested in. Lots of work remains, e.g.,
 How the feature should be modeled in a simple way, if PLDA does not work? How the neural-learned features
 generalize from one task (or language) to another? How to involve auxiliary information to boost the model
 quality? All need careful investigation.

\bibliographystyle{IEEEtran}
\bibliography{dvector}

\begin{thebibliography}{10}
\providecommand{\url}[1]{#1}
\csname url@samestyle\endcsname
\providecommand{\newblock}{\relax}
\providecommand{\bibinfo}[2]{#2}
\providecommand{\BIBentrySTDinterwordspacing}{\spaceskip=0pt\relax}
\providecommand{\BIBentryALTinterwordstretchfactor}{4}
\providecommand{\BIBentryALTinterwordspacing}{\spaceskip=\fontdimen2\font plus
\BIBentryALTinterwordstretchfactor\fontdimen3\font minus
  \fontdimen4\font\relax}
\providecommand{\BIBforeignlanguage}[2]{{%
\expandafter\ifx\csname l@#1\endcsname\relax
\typeout{** WARNING: IEEEtran.bst: No hyphenation pattern has been}%
\typeout{** loaded for the language `#1'. Using the pattern for}%
\typeout{** the default language instead.}%
\else
\language=\csname l@#1\endcsname
\fi
#2}}
\providecommand{\BIBdecl}{\relax}
\BIBdecl

\bibitem{Reynolds00}
D.~Reynolds, T.~Quatieri, and R.~Dunn, ``Speaker verification using adapted
  gaussian mixture models,'' \emph{Digital Signal Processing}, vol.~10, no.~1,
  pp. 19--41, 2000.

\bibitem{Kenny07}
P.~Kenny, G.~Boulianne, P.~Ouellet, and P.~Dumouchel, ``Joint factor analysis
  versus eigenchannels in speaker recognition,'' \emph{IEEE Transactions on
  Audio, Speech, and Language Processing}, vol.~15, pp. 1435--1447, 2007.

\bibitem{dehak2011front}
N.~Dehak, P.~J. Kenny, R.~Dehak, P.~Dumouchel, and P.~Ouellet, ``Front-end
  factor analysis for speaker verification,'' \emph{IEEE Transactions on Audio,
  Speech, and Language Processing}, vol.~19, no.~4, pp. 788--798, 2011.

\bibitem{Campbell06}
W.~Campbell, D.~Sturim, and D.~Reynolds, ``Support vector machines using gmm
  supervectors for speaker verification,'' \emph{Signal Processing Letters,
  IEEE}, vol.~13, no.~5, pp. 308--311, 2006.

\bibitem{Ioffe06}
S.~Ioffe, ``Probabilistic linear discriminant analysis,'' \emph{Computer Vision
  ECCV 2006, Springer Berlin Heidelberg}, pp. 531--542, 2006.

\bibitem{Kenny14}
P.~Kenny, V.~Gupta, T.~Stafylakis, P.~Ouellet, and J.~Alam, ``Deep neural
  networks for extracting baum-welch statistics for speaker recognition,''
  \emph{Odyssey}, 2014.

\bibitem{lei2014novel}
Y.~Lei, N.~Scheffer, L.~Ferrer, and M.~McLaren, ``A novel scheme for speaker
  recognition using a phonetically-aware deep neural network,'' in
  \emph{Acoustics, Speech and Signal Processing (ICASSP), 2014 IEEE
  International Conference on}.\hskip 1em plus 0.5em minus 0.4em\relax IEEE,
  2014, pp. 1695--1699.

\bibitem{Kinnunen10}
T.~Kinnunen and H.~Li, ``An overview of text-independent speaker recognition:
  From features to supervectors,'' \emph{Speech communication}, vol.~52, no.~1,
  pp. 12--40, 2010.

\bibitem{hinton2012deep}
G.~Hinton, L.~Deng, D.~Yu, G.~E. Dahl, A.-r. Mohamed, N.~Jaitly, A.~Senior,
  V.~Vanhoucke, P.~Nguyen, T.~N. Sainath \emph{et~al.}, ``Deep neural networks
  for acoustic modeling in speech recognition: The shared views of four
  research groups,'' \emph{IEEE Signal Processing Magazine}, vol.~29, no.~6,
  pp. 82--97, 2012.

\bibitem{ehsan14}
V.~Ehsan, L.~Xin, M.~Erik, L.~M. Ignacio, and G.-D. Javier, ``Deep neural
  networks for small footprint text-dependent speaker verification,'' in
  \emph{Acoustics, Speech and Signal Processing (ICASSP), 2014 IEEE
  International Conference on}, vol.~28, no.~4, 2014, pp. 357--366.

\bibitem{heigold2016end}
G.~Heigold, I.~Moreno, S.~Bengio, and N.~Shazeer, ``End-to-end text-dependent
  speaker verification,'' in \emph{Acoustics, Speech and Signal Processing
  (ICASSP), 2016 IEEE International Conference on}.\hskip 1em plus 0.5em minus
  0.4em\relax IEEE, 2016, pp. 5115--5119.

\bibitem{zhang2017end}
S.-X. Zhang, Z.~Chen, Y.~Zhao, J.~Li, and Y.~Gong, ``End-to-end attention based
  text-dependent speaker verification,'' \emph{arXiv preprint
  arXiv:1701.00562}, 2017.

\bibitem{liu2015deep}
Y.~Liu, Y.~Qian, N.~Chen, T.~Fu, Y.~Zhang, and K.~Yu, ``Deep feature for
  text-dependent speaker verification,'' \emph{Speech Communication}, vol.~73,
  pp. 1--13, 2015.

\bibitem{snyderdeep16}
D.~Snyder, P.~Ghahremani, D.~Povey, D.~Garcia-Romero, Y.~Carmiel, and
  S.~Khudanpur, ``Deep neural network-based speaker embeddings for end-to-end
  speaker verification,'' in \emph{SLT'2016}, 2016.

\bibitem{yao2016speaker}
T.~Yao, C.~Meng, H.~Liang, and L.~Jia, ``Speaker recognition system based on
  deep neural networks and bottleneck features,'' \emph{Journal of Tsinghua
  University (Science and Technology)}, vol.~56, no.~11, pp. 1143--1148, 2016.

\bibitem{povey2014parallel}
D.~Povey, X.~Zhang, and S.~Khudanpur, ``Parallel training of dnns with natural
  gradient and parameter averaging,'' \emph{arXiv preprint arXiv:1410.7455},
  2014.

\bibitem{li2015improved}
L.~Li, Y.~Lin, Z.~Zhang, and D.~Wang, ``Improved deep speaker feature learning
  for text-dependent speaker recognition,'' in \emph{Signal and Information
  Processing Association Annual Summit and Conference (APSIPA), 2015
  Asia-Pacific}.\hskip 1em plus 0.5em minus 0.4em\relax IEEE, 2015, pp.
  426--429.

\bibitem{povey2011kaldi}
D.~Povey, A.~Ghoshal, G.~Boulianne, L.~Burget, O.~Glembek, N.~Goel,
  M.~Hannemann, P.~Motlicek, Y.~Qian, P.~Schwarz \emph{et~al.}, ``The kaldi
  speech recognition toolkit,'' in \emph{IEEE 2011 workshop on automatic speech
  recognition and understanding}, no. EPFL-CONF-192584.\hskip 1em plus 0.5em
  minus 0.4em\relax IEEE Signal Processing Society, 2011.

\bibitem{saaten2008}
L.~v.~d. Maaten and G.~Hinton, ``Visualizing data using t-sne,'' \emph{Machine
  Learning Research}, 2008.

\end{thebibliography}

\end{document}